\documentclass[prl,twocolumn,showpacs,superscriptaddress,floatfix]{revtex4}  

\usepackage{graphicx}
\usepackage{dcolumn}
\usepackage{bm}
\usepackage[mathlines]{lineno}
\usepackage{mathtools}

\DeclarePairedDelimiter \abs{\lvert}{\rvert}

\begin{document}

\preprint{APS/123-QED}

\title{Antiferromagnetic domain wall motion driven by spin-orbit torques}

\author{Takayuki Shiino}
\thanks{These two authors contributed equally to this work.}
\affiliation{Department of Materials Science and Engineering, KAIST, Daejeon, 305-701, Korea}

\author{Se-Hyeok Oh}
\thanks{These two authors contributed equally to this work.}
\affiliation{Department of Nano-Semiconductor and Engineering, Korea University, Seoul 136-701, Korea}

\author{Paul M. Haney}
\affiliation{Center for Nanoscale Science and Technology, National Institute of Standards and Technology, Gaithersburg, Maryland 20899-6202, USA}

\author{Seo-Won Lee}
\affiliation{Department of Materials Science and Engineering, Korea University, Seoul 136-701, Korea}

\author{Gyungchoon Go}
\affiliation{Department of Materials Science and Engineering, Korea University, Seoul 136-701, Korea}

\author{Byong-Guk Park}
\email{bgpark@kaist.ac.kr}
\affiliation{Department of Materials Science and Engineering, KAIST, Daejeon, 305-701, Korea}

\author{Kyung-Jin Lee}
\email{kj_lee@korea.ac.kr}
\affiliation{Department of Nano-Semiconductor and Engineering, Korea University, Seoul 136-701, Korea}
\affiliation{Department of Materials Science and Engineering, Korea University, Seoul 136-701, Korea}
\affiliation{KU-KIST Graduate School of Converging Science and Technology, Korea University, Seoul 136-701, Korea}

\date{\today}

\begin{abstract}
We theoretically investigate dynamics of antiferromagnetic domain walls driven by spin-orbit torques in antiferromagnet/heavy metal bilayers. We show that spin-orbit torques drive antiferromagnetic domain walls much faster than ferromagnetic domain walls. As the domain wall velocity approaches the maximum spin-wave group velocity, the domain wall undergoes Lorentz contraction and emits spin-waves in the terahertz frequency range. The interplay between spin orbit torques and the relativistic dynamics of antiferromagnetic domain walls leads to the efficient manipulation of antiferromagnetic spin textures and paves the way for the generation of high frequency signals in antiferromagnets.
\end{abstract}


\maketitle



Antiferromagnets are ordered spin systems in which the magnetic moments are compensated on an atomic scale. The antiferromagnetic order and consequent zero net magnetic moment are maintained by antiferromagnetic exchange coupling of neighboring spins. Any external disturbance competes directly with the large antiferromagnetic exchange, which results in magnetic excitations in terahertz frequency ranges~\cite{Kampfrath2011}. Furthermore, an antiferromagnet has no magnetic stray field, which is beneficial for integrated circuits because the stray field is a primary source of detrimental magnetic perturbations~\cite{MacDonald2011,Duine2011}. These attractive features of antiferromagnets have led to the recent development of {\it antiferromagnetic spintronics}, an emerging research field which pursues the use of antiferromagnets as active elements in spintronic-based devices~\cite{Jungwirth2016}.

The principal discipline of antiferromagnetic spintronics is the robust detection and manipulation of the antiferromagnetic order. The antiferromagnetic order can be electrically probed through the (tunneling) anisotropic magnetoresistance effect~\cite{TAMR} or the spin pumping effect~\cite{Cheng2014}. Significant progress has also been made on the manipulation of the antiferromagnetic order, using both charge and spin currents~\cite{EarlySTT}. Conventional spin-transfer torque enables current-driven manipulation of antiferromagnetic spin textures such as antiferromagnetic domain walls~\cite{Swaving2011,Hals2011,Tveten2013} and antiferromagnetic skyrmions~\cite{Zhang2015arXiv,Barker2015arXiv}. We note however that most previous studies on current-driven manipulation of antiferromagnetic order have neglected spin-orbit coupling.

\begin{figure}[t]
\begin{center}
\includegraphics[scale=0.20]{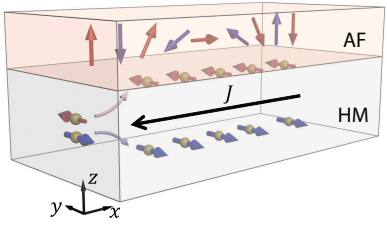}
\caption{
Schematic illustration of an antiferromagnet (AF)/heavy metal (HM) bilayer system. An in-plane charge current $J$ generates a perpendicular spin current, which in turn generates SOTs acting on antiferromagnetic moments.
}
\label{fig:FIG1}
\end{center}
\end{figure}

The influence of spin-orbit coupling on spin transport and magnetization dynamics has recently attracted considerable attention, as it enables the study of fundamental interactions among conduction electron spin, electron orbit, and local magnetization. In ferromagnet/heavy metal bilayers, an in-plane current generates spin-orbit spin-transfer torques (SOTs)~\cite{SOTexp,SOTtheory}. The microscopic origin of these torques remain under debate, but they can be classified according to their direction.  In the coordinate system of Fig. \ref{fig:FIG1}, the ``field-like" torque induces precession of spins around the $y$-axis, while the ``damping-like" torque directs the spin towards the $y$-axis.  Spin-orbit coupling additionally induces a noncollinear magnetic exchange in these bilayer systems known as the interfacial Dzyaloshinskii-Moriya interaction (DMI), which stabilizes N{\'e}el domain walls in ferromagnets. The SOT combined with DMI efficiently drives a ferromagnetic domain wall~\cite{Thiaville2012,SOTDWexp}. Recently, current-driven relativistic N{\'e}el-order fields in antiferromagnets have been predicted theoretically~\cite{Zelezny2014} and confirmed experimentally~\cite{Wadley2016}, indicating the relevance of SOT in antiferromagnets with inversion asymmetry. This relativistic N{\'e}el-order field is present in only a specific class of antiferromagnets with the inversion asymmetry of antiferromagnetic order.

In this Letter, we investigate SOT-driven antiferromagnetic domain wall motion in antiferromagnet/heavy metal bilayers in the presence of interfacial DMI, based on the collective coordinate approach~\cite{Swaving2011,Hals2011,Tveten2013} and atomistic spin model simulations~\cite{Evans2014}. Because SOTs in antiferromagnet/heavy metal bilayers emerge by the {\it structural inversion asymmetry}, our result is applicable to a wide variety of antiferromagnets in contact with a heavy metal layer. We show that at reasonable current densities the antiferromagnetic domain wall velocity can reach a few kilometers per second, which is much larger than that of ferromagnetic domain wall. As the wall velocity approaches the maximum group velocity of spin-waves, it undergoes Lorentz contraction and emits spin-waves with wavelength on the order of the material lattice constant. The frequency of emitted spin waves is in the terahertz range and thus the antiferromagnetic domain wall can be used as a direct-current-driven terahertz source.

We consider an antiferromagnetic domain wall in a one-dimensional nanowire system composed of antiferromagnet/heavy metal bilayer with perpendicular magnetic anisotropy (Fig. \ref{fig:FIG1}). An in-plane current flowing along the $x$-axis generates field-like and damping-like SOTs~\cite{SOTtheory}. For the analytical description, we use the nonlinear sigma model in the continuum approximation~\cite{Hals2011}. To begin, we define the total and staggered magnetization as follows:
$\mathbf{m} \equiv \mathbf{m}_{1}(x,t) + \mathbf{m}_{2}(x,t)$
and
$\mathbf{l} \equiv \mathbf{m}_{1}(x,t) - \mathbf{m}_{2}(x,t)$
where
$\mathbf{m}_{1}(x,t)$
and
$\mathbf{m}_{2}(x,t)$
are respectively the magnetic moment densities of two sub-lattices with
$\abs{\mathbf{m}_{1}(x,t)} = \abs{\mathbf{m}_{2}(x,t)} = m_\text{s}$.
The constraints are
$\mathbf{m} \cdot \mathbf{l} = 0$
and
$\mathbf{m}^2 + \mathbf{l}^2 = l^2$
where
$ l = 2m_\text{s} $.
In the following, we discuss the antiferromagnetic domain wall dynamics with $\mathbf{m}(x,t)$ and $\mathbf{n}(x,t) (\equiv \mathbf{l}(x,t) / l) $ and expand equations up to second order in small parameters~\cite{Swaving2011}, assuming that time-, space-derivative, damping, SOTs, anisotropy, and interfacial DMI are small.

The leading-order free energy in the continuum approximation is
$ U = \int [
\frac{a}{2} \abs{\mathbf{m}}^2
+   \frac{A}{2} ( \frac{\partial \mathbf{n}}{\partial x } )^2
+   L \mathbf{m} \cdot \frac{\partial \mathbf{n}}{\partial x }
-   \frac{K}{2} (   \mathbf{e}_{z}  \cdot   \mathbf{n}   )^2
+   \frac{D}{2}  \mathbf{e}_{y}   \cdot   (  \mathbf{n}    \times    \frac{\partial \mathbf{n}}{\partial x }   )
]  d\mathbf{r}  $,
where $a$ and $A$ are the homogeneous and inhomogeneous exchange constants, respectively, $L$ is the parity-breaking exchange constant~\cite{Papanicolau1995,Tveten2016}, and $K$ and $D$ denote the easy-axis anisotropy and interfacial DMI, respectively. The third term breaks the parity of the energy and expresses the intrinsic magnetization of the texture~\cite{Papanicolau1995,Tveten2016}.
Here $a$ and $A$ are related by
$ A = a l^2 {\Delta}^2 / 8 $, $a$ and $L$ by $ L = a l \Delta / 4 $, $\Delta$ ($=2d$) is the length of the antiferromagnetic unit cell, and $d$ is the lattice constant.
From the functional derivative of the energy density, we obtain the lowest order in the effective fields
$ \mathbf{f}_\text{m} = - \frac{\delta U}{\delta \mathbf{m}}
= - a \mathbf{m} - L \frac{\partial \mathbf{n}}{\partial x }   $
and
$ \mathbf{f}_\text{n} = - \frac{\delta U}{\delta \mathbf{n}}
= A \frac{{\partial}^2 \mathbf{n}}{\partial x^2 }
+ L \frac{\partial \mathbf{m}}{\partial x }
+ K (   \mathbf{e}_{z}  \cdot   \mathbf{n}   ) \mathbf{e}_{z}
+ D ( \mathbf{e}_{y}  \times  \frac{\partial \mathbf{n}}{\partial x }   )
$.

Disregarding nonlinear terms, the equations of motion are:
\begin{eqnarray}
\frac{\partial \mathbf{n}}{\partial t }
& = &
( \gamma \mathbf{f}_\text{m} - G_1  \frac{\partial \mathbf{m}}{\partial t }   ) \times \mathbf{n} + \mathbf{T}^{\text{n}}_\text{SOT}, \label{Eq1} \\
\frac{\partial \mathbf{m}}{\partial t }
& = &
( \gamma \mathbf{f}_\text{n} - G_2  \frac{\partial \mathbf{n}}{\partial t }   ) \times \mathbf{n} + \mathbf{T}^{\text{m}}_\text{SOT}, \label{Eq2}
\end{eqnarray}
where $\gamma$ is the gyromagnetic ratio, and $G_1$ and $G_2$ are damping parameters~\cite{Hals2011,Tveten2013}. Rewriting the field-like and damping-like torques in terms of $\mathbf{n}$ and $\mathbf{m}$ and retaining lowest order terms leads to:
$ \mathbf{T}^{\text{n}}_\text{SOT}
= \frac{\gamma B_\text{D}}{l} \mathbf{n} \times ( \mathbf{m} \times \mathbf{e}_y )
+ \gamma B_\text{F} \mathbf{n} \times \mathbf{e}_y $
and
$ \mathbf{T}^{\text{m}}_\text{SOT}
= \gamma B_\text{D} l \mathbf{n} \times ( \mathbf{n} \times \mathbf{e}_y )
+ \gamma B_\text{F} \mathbf{m} \times \mathbf{e}_y
$~\cite{Cheng2014}
where $B_\text{D} (=\mu_\text{B} \theta_\text{SH} J /\gamma e m_\text{s} t_z)$ and $B_\text{F} (= \chi B_\text{D})$ denote effective fields corresponding to the damping-like and field-like components of SOT, respectively, $t_z$ is the thickness of antiferromagnet, $\theta_\text{SH}$ is the effective spin Hall angle, $\mu_\text{B}$ is the Bohr magneton, $e$ is the electron charge, $J$ is the current density, and $\chi$ is the ratio of $B_\text{F}$ to $B_\text{D}$.


We introduce the collective coordinates for the domain wall position $r$ and angle $\phi$, and the Walker ansatz for the wall profile~\cite{Walker1974,Tveten2013}:
$ \mathbf{n}(x,t) = (n_x,n_y,n_z)
= (\sin \theta \cos \phi, \sin \theta \sin \phi, \cos \theta )$
where $\theta = 2 \tan^{-1}[\exp (\frac{x-r}{\lambda})]$, and $\lambda$ is the domain wall width. Following the procedure in Ref.~\cite{Tveten2013}, $\mathbf{m}$ can be expressed in terms of $\mathbf{n}$ by combining Eqs. (\ref{Eq1}) and (\ref{Eq2}). Substituting the Walker profile into $\mathbf{n}$ and keeping leading order terms, we obtain the following equations:
\begin{eqnarray}
\ddot{r} &+& a \gamma G_2 \dot{r}
+ \frac{\pi}{2} a \gamma^2 l \lambda B_\text{D} \cos \phi
+ \frac{3\pi}{4} \gamma \lambda B_\text{F} \dot{\phi} \sin \phi  =  0, \label{Eq3} \\
\ddot{\phi} &+& a \gamma G_2 \dot{\phi}
- \frac{\pi }{4} \frac{a \gamma^2 }{\lambda} D \sin \phi
- \frac{\pi }{2 } \frac{\gamma }{\lambda} B_\text{F} \dot{r} \sin \phi  =  0.  \label{Eq4}
\end{eqnarray}

We first consider the case for a N{\'e}el wall (i.e., $\phi(t=0)=0$ or $\pi$), which is stabilized by nonzero $D$ since the hard-axis anisotropy of antiferromagnetic domain wall is negligible. In Eqs. (\ref{Eq3}) and (\ref{Eq4}), all terms having $\sin\phi$ are zero at $t=0$. With $\dot r=0$ and $\dot \phi=0$ at $t=0$ (i.e., the domain wall is at rest at $t=0$), $\dot \phi$ is always zero and the steady-state velocity $v_\text{DW}$ of N{\'e}el wall is given as
\begin{equation}
v_\text{DW} = v_\text{AF} = -\pi \gamma \lambda B_\text{D}/2\alpha, \label{Eq5}
\end{equation}
where $\alpha$ ($=G_2/l$) is the Gilbert damping. It is worthwhile comparing $v_\text{AF}$ to the velocity $v_\text{F}$ of a N{\'e}el type ferromagnetic domain wall driven by SOT ($v_\text{F}=\gamma \pi D/(2 m_\text{s} \sqrt{1+(\alpha D/B_\text{D} m_\text{s} \lambda)^2})$~\cite{Thiaville2012}). In the small $B_\text{D}$ limit, $|v_\text{F}|=|v_\text{AF}|$. This equivalence is however broken when $B_\text{D}$ is large. For a ferromagnetic wall, $\phi$ increases with $B_\text{D}$ so that $v_\text{F}$ saturates to $\gamma \pi D/2 m_\text{s}$. For an antiferromagnetic wall, on the other hand, $\phi$ does not vary with time and as a result, $v_\text{AF}$ increases linearly with $B_\text{D}$ (thus $J$). This unique property of antiferromagnetic N\'{e}el wall leads to a large $v_\text{AF}$ especially for a small damping $\alpha$ because $v_\text{AF} \propto 1/\alpha$. A small damping is realized in semiconducting or insulating antiferromagnets where spin scatterings are suppressed.


We next consider the case for a Bloch wall (i.e., $\phi(t=0)=\pi/2$ or $3\pi/2$), corresponding to $D=0$. From Eq. (\ref{Eq4}), $\dot \phi$ is always zero because $\dot r=0$ and $\dot \phi=0$ at $t=0$. Substituting $\dot \phi=0$ and $\cos\phi=0$ in Eq. (\ref{Eq3}), we find $v_\text{DW}$ of a Bloch wall is zero when it is driven only by the SOT.

To verify the analytical results, we perform numerical calculations with the atomistic Landau-Lifshitz-Gilbert (LLG) equation \cite{Evans2014}. The Hamiltonian of antiferromagnets is
$
{\cal H} = A_\text{sim} \Sigma_{i} \mathbf{S}_{i} \cdot \mathbf{S}_{i+1}
- K_\text{sim} \Sigma_{i}  ( \mathbf{S}_{i} \cdot \mathbf{e}_\text{z}   )^2
- D_\text{sim} \Sigma_{i} \mathbf{e}_z \cdot ( \mathbf{S}_{i} \times \mathbf{S}_{i+1} )
+ \frac{\mu_0}{8\pi} m_\text{s} \mu \Sigma_{i,j}  ( \mathbf{S}_{i} \cdot \mathbf{S}_{j}  - \frac{3(\mathbf{S}_{i} \cdot \mathbf{r}_{ij})(\mathbf{S}_{j} \cdot \mathbf{r}_{ij})}{r^2} )
$,
where $\mathbf{S}_{i}$ represents the normalized magnetic moment (i.e., $\abs{\mathbf{S}_{i}}=1$) at lattice site $i$, $\mu$ is the magnetic moment per lattice site, and $A_\text{sim}, K_\text{sim}, D_\text{sim}$
denote the exchange, anisotropy, and DMI energies, respectively.
The last term represents dipole-dipole interaction where $\mathbf{r}_{ij}$ is a distance vector between lattice sites $i$ and $j$ (i.e., $\abs{\mathbf{r}_{ij}}=r$).
The atomistic LLG equation is as follows:
$
\frac{\partial \mathbf{S}_{i}}{\partial t}
= - \gamma \mathbf{S}_{i} \times \mathbf{B}_\text{eff}
+ \alpha \mathbf{S}_{i} \times  \frac{\partial \mathbf{S}_{i}}{\partial t}
+ \gamma B_\text{D}  \mathbf{S}_{i} \times ( \mathbf{S}_{i} \times \mathbf{e}_y )
+ \gamma B_\text{F}   ( \mathbf{S}_{i} \times \mathbf{e}_y )
$,
where $\mathbf{B}_\text{eff} = - \frac{1}{\mu}\frac{\delta {\cal H}}{\delta \mathbf{S}_{i}}$
is the effective field.
We use modeling parameters as follows~\cite{Archer2011}:
$
d = 0.4\, \mathrm{nm}, A_\text{sim} = 16.0\, \mathrm{meV} ,
K_\text{sim} = 0.04\, \mathrm{meV} , \mu = 3.45 \mu_\text{B}
$, $\theta_\text{SH}=0.1$, $\alpha = 0.001$, and $\chi = 0$ (i.e., $B_\text{F}=0$) or 23 (i.e., $B_\text{F} \ne 0$~\cite{Aurelien}).
We use $D_\text{sim} = 0$ or $D_\text{sim} = 2.0 \, \mathrm{meV}$, obtaining a Bloch or N\'{e}el wall, respectively.

\begin{figure}[tb]
\begin{center}
\includegraphics[scale=0.47]{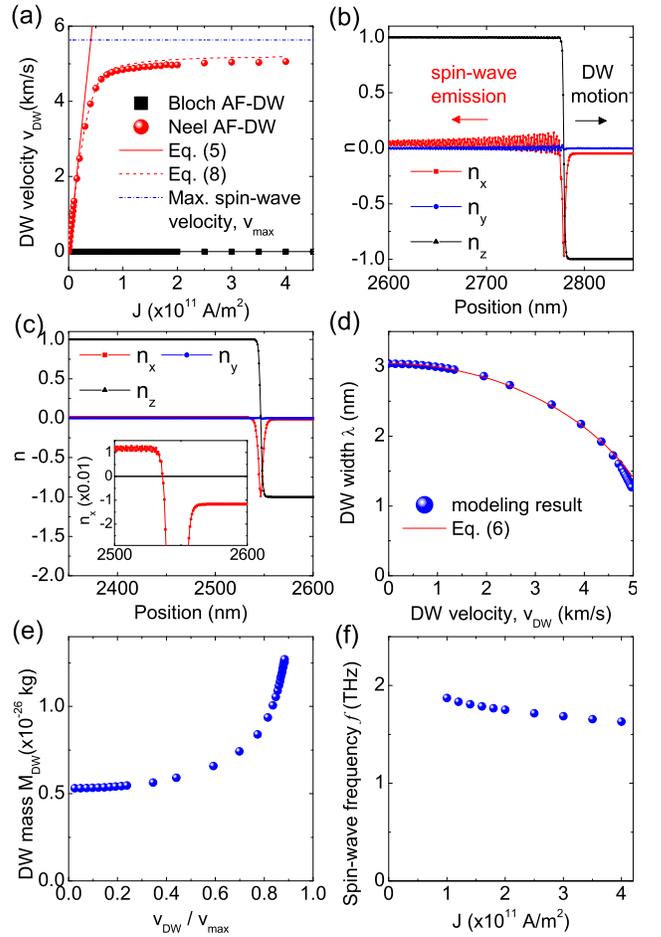}
\caption{
SOT-driven antiferromagnetic domain wall motion for $B_\text{F}=0$: (a) Domain wall velocity $v_\text{DW}$ vs current density $J$~\cite{note}. (b) Configuration of N\'{e}el-type antiferromagnetic domain wall during the steady motion at $J=0.5 \times 10^{11} \, \mathrm{A/m^2}$. (c) Configuration of equilibrium N\'{e}el-type antiferromagnetic domain wall. Inset shows $n_x$ component. (d) Domain wall width $\lambda$ vs domain wall velocity $v_\text{DW}$. (e) Domain wall mass $M_\text{DW}$ vs $v_\text{DW}/v_\text{max}$ where $v_\text{max}$ is the maximum group velocity of spin-wave. (f) Spin-wave frequency $f$ vs $J$.
}
\label{fig:FIG2}
\end{center}
\end{figure}

The symbols in Fig. \ref{fig:FIG2}(a) show numerical results of the steady-state velocity $v_\text{DW}$ as a function of the current density $J$ when $B_\text{F}=0$. As predicted by Eq.~(\ref{Eq5}), a Bloch wall does not move whereas the N\'{e}el wall velocity linearly increases with $J$ in a low current regime. We find however that the N\'{e}el wall velocity saturates in a high current regime, in contrast to the prediction of Eq.~(\ref{Eq5}). As explained above, such saturation behavior of $v_\text{DW}$ is also expected for a ferromagnetic wall when it is driven by combined effects of SOT and DMI~\cite{Thiaville2012}. In case of ferromagnetic walls, the saturation of $v_\text{DW}$ results from the saturation of the domain wall angle $\phi$ in the high current regime. In case of antiferromagnetic walls, however, $\phi$ does not change with time [i.e., $\dot \phi=0$; see Eq.~(\ref{Eq4}) and Fig. \ref{fig:FIG2}(b)] so that the $v_\text{DW}$ saturation of an antiferromagnetic domain wall results from a completely different origin.

We find that the spin-wave emission from the antiferromagnetic domain wall is the origin of the $v_\text{DW}$ saturation in a high current regime. A snap-shot configuration of $\mathbf{n}$ shows that the wall moves to the right while emitting spin-waves to the left [Fig. \ref{fig:FIG2}(b); see supplementary movie in~\cite{SOM}]. The reason for spin-wave emission is as follows: The damping-like SOT asymmetrically tilts the domains on the right and the left of wall [see inset of Fig. \ref{fig:FIG2}(c)]. Because of the asymmetric domain tilting, the rear (i.e., left) of wall has a steeper gradient of $\mathbf{n}$ and thus a higher exchange energy than the front of wall. As the wall moves faster, the wall width $\lambda$ shrinks more [see Fig. \ref{fig:FIG2}(d)]. As $\lambda$ approaches the lattice constant, the antiferromagnetic domain wall is unable to sustain its energy and starts to emit spin-waves towards its rear (where the gradient is steeper) to release the energy. Therefore, the spin-wave emission serves as an additional energy dissipation channel and slows down the wall motion.

This interesting dynamics of antiferromagnetic domain walls in the high current regime is a manifestation of the relativistic kinematics originating from the Lorentz invariance of the magnon dispersion~\cite{Haldane1983,KimSeKwon2014}. In special relativity, as the velocity of a massive particle approaches the speed of light $c$, it shrinks via Lorentz contraction and its velocity saturates to $c$. For the dynamics of antiferromagnets, the speed of light is replaced by the maximum spin-wave group velocity because the antiferromagnetic domain wall can be decomposed into spin-waves and has a finite inertial mass~\cite{KimSeKwon2014}. The velocity limit of an antiferromagnetic domain wall can therefore be described by the relativistic kinematics: it undergoes Lorentz contraction as its velocity approaches the maximum spin-wave group velocity, and its velocity saturates to the maximum spin-wave group velocity. Figure \ref{fig:FIG2}(d) shows that numerically obtained $\lambda$ indeed shrinks as $v_\text{DW}$ becomes larger. The Lorentz contraction of antiferromagnetic domain wall is described by
\begin{equation}
\lambda = \lambda_\text{eq} \sqrt{1-(v_\text{DW}/v_\text{max})^2}, \label{Eq6}
\end{equation}
where $\lambda_\text{eq}$ is the equilibrium domain wall width and $v_\text{max}$ is the maximum group velocity of spin-wave. To obtain $v_\text{max}$, we consider spin-waves in the bulk domain regions for simplicity. Spin-waves are set as
$\mathbf{n} = [1-(n_x^2 + n_y^2)/2]\mathbf{e}_z
+ n_x \mathbf{e}_x
+ n_y \mathbf{e}_y $
and
$\mathbf{m} = - (m_x n_x + m_y n_y) \mathbf{e}_z
+ m_x \mathbf{e}_x
+ m_y \mathbf{e}_y $,
which satisfies the constraints $\abs{\mathbf{n}} = 1$ and $\mathbf{n} \cdot \mathbf{m} = 0$ within the second order~\cite{Tveten2014}. Inserting $\mathbf{n}$ and $\mathbf{m}$ in Eqs. (\ref{Eq1}) and (\ref{Eq2}), expanding to the first order in $n_x$ and $n_y$, and neglecting the dissipation term, we obtain a wave equation for $n_x$ as follows:
\begin{equation}
\frac{\partial^2 n_x}{\partial t^2}
=
a \gamma^2 \tilde{A} \frac{\partial^2 n_x}{\partial x^2}
- a \gamma^2 K n_x
\pm a \gamma^2 l B_\text{D},
\label{Eq7}
\end{equation}
where $\tilde{A}=A-L^2/a$ and the upper (lower) sign corresponds to the up (down) domain. A plane wave solution of Eq. (\ref{Eq7}) is
$ n_x = \delta n_x \exp[i(k x - \omega t)] \pm \frac{l B_\text{D}}{K}$,
where $\delta n_x$ is the spin-wave amplitude and the second term describes the domain tilting induced by damping-like SOT [inset of Fig. \ref{fig:FIG2}(c)]. The dispersion relation is then given by
$ \omega = \gamma \sqrt{a(\tilde{A}k^2 + K)} $,
from which we obtain the group velocity
$
v_\text{g} = \frac{d\omega}{dk} = \gamma a l d / (2\sqrt{1+\frac{4K}{a l^2 d^2 k^2}})
$, and thus $v_\text{max} = \gamma a l d / 2$. For the modeling parameters, $v_\text{max}$ is about 5.6 km/s as shown in Fig. \ref{fig:FIG2}(a). With $v_\text{max}$ given above, the relativistically corrected $v_\text{DW}$ is given as
\begin{equation}
v_\text{DW} = \frac{\gamma a l d}{2} \sqrt{1-(\lambda/\lambda_\text{eq})^2}. \label{Eq8}
\end{equation}
Equations (\ref{Eq6}) and (\ref{Eq8}) describe the numerical results reasonably well [see Fig. \ref{fig:FIG2}(a) and (d)].

Two remarks on the relativistic kinematics of SOT-driven antiferromagnetic domain wall motion are in order. Firstly, it is also associated with the inertial mass of the wall. In steady-state motion,
the effective inertial mass $M_\text{DW}$ of antiferromagnetic domain wall is given by $M_\text{DW} = \rho \int \frac{\partial \mathbf{n}}{\partial r} \cdot \frac{\partial \mathbf{n}}{\partial r} dV$ where $\rho = 1/a\gamma^2$~\cite{KimSeKwon2014}.
With the Walker ansatz, $M_\text{DW}$ is represented as
$ M_\text{DW} = 2\rho w t_z / \lambda
= 2\rho w t_z / \lambda_\text{eq} \sqrt{1-(v_\text{DW}/v_\text{max})^2}$
where $w$ is the wire width. Because of the Lorentz contraction, $M_\text{DW}$ increases by the Lorentz factor $1/\sqrt{1-(v_\text{DW}/v_\text{max})^2}$ as $v_\text{DW}$ increases (Fig. \ref{fig:FIG2}(e)). Secondly, the frequency of emitted spin-waves is in the terahertz range. Using the modeling parameters in the spin-wave dispersion given above, one finds that the spin-wave frequency $f_\text{max}$ (=$\omega/2\pi$) corresponding to $v_\text{max}$ is about 2.5 THz. The numerically obtained spin-wave frequency is slightly lower than $f_\text{max}$ but is still in the terahertz range [Fig. \ref{fig:FIG2}(f)]. This suggests that the antiferromagnetic domain wall can be used as a terahertz source of electric signal.

\begin{figure}[tb]
\begin{center}
\includegraphics[scale=0.47]{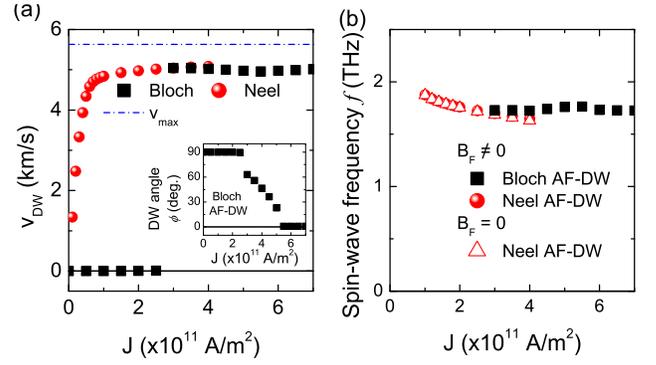}
\caption{
SOT-driven antiferromagnetic domain wall motion for $B_\text{F} \ne 0$ ($\chi=23$~\cite{Aurelien}): (a) Domain wall velocity $v_\text{DW}$ vs current density $J$~\cite{note}. Inset shows the domain wall angle $\phi$ for an antiferromagnetic domain wall that is initially of Bloch type. (b) Spin-wave frequency $f$ vs $J$. $f$ for $B_\text{F}=0$ is also shown for comparison.
}
\label{fig:FIG3}
\end{center}
\end{figure}

We next show numerical results for $B_\text{F}\ne 0$ (Fig. \ref{fig:FIG3}). $B_\text{F}$ does not affect dynamics of the N\'{e}el wall: $v_\text{DW}$ of the N\'{e}el wall is almost independent of $B_\text{F}$. On the other hand, $B_\text{F}$ affects dynamics of Bloch wall substantially. For $B_\text{F}=0$ the Bloch wall does not move [Fig. \ref{fig:FIG2}(a)] whereas for $B_\text{F} \ne 0$ it moves with $v_\text{DW} \approx v_\text{max}$ above a certain threshold current density [$J_\text{th}=2.5 \times 10^{11} \, \mathrm{A/m^2}$; see Fig. \ref{fig:FIG3}(a)]. This fast motion of the Bloch wall is accompanied by a current-dependent change in the domain wall angle $\phi$ [inset of Fig. \ref{fig:FIG3}(a)], because a nonzero $B_\text{F}$ transforms an initial Bloch wall into a N\'{e}el type wall. This transformation is known as the {\it spin-flop} transition of an antiferromagnet~\cite{Morrish}. When an antiferromagnet is subject to a large magnetic field applied along the staggered magnetization $\mathbf{n}$, the spin sublattice antiparallel to the applied field is energetically unfavorable. At a threshold field, the spins {\it flop} to a configuration where both sublattices are perpendicular to the applied field~\cite{Nogues}, which corresponds to the transformation from a Bloch to a N\'{e}el wall. From Fig. \ref{fig:FIG3}(a), we find that $v_\text{DW}$ saturates in the high current regime as in the case with $B_\text{F}=0$. This $v_\text{DW}$ saturation also originates from the emission of spin-waves in the terahertz frequency ranges [Fig. \ref{fig:FIG3}(b)].

To conclude, the SOT can efficiently move the antiferromagnetic domain wall. The damping-like SOT is the main driving force whereas the field-like SOT is effective by transforming a Bloch wall into a N\'{e}el wall. The antiferromagnetic domain wall velocity can reach a few kilometers per second, which is orders of magnitude larger than the ferromagnetic domain wall velocity. The relativistic kinematics of antiferromagnets results in the saturation of $v_\text{DW}$ in the high current regime, which is accompanied by the emission of spin-waves with frequency in the terahertz range. An antiferromagnetic domain wall can therefore serve as a terahertz source. We finally note that the relativistic kinematics is not unique to antiferromagnetic domain walls: a ferromagnetic domain wall can exhibit relativistic motion in systems with biaxial anisotropy, which is essential for a finite inertial mass. Wang {\it et al.}~\cite{WangXS2014} reported field-driven ferromagnetic domain wall motion with spin-waves emission. This relativistic motion is however realized only by assuming very large hard-axis anisotropy, comparable to exchange energy. This unrealistic assumption is required to push the wall width to a few lattice constants. In contrast, for antiferromagnetic domain walls, the condition of a-few-lattice-constant wall width is naturally realized by the SOT.

\begin{acknowledgments}
We acknowledge fruitful discussions with A. Manchon, J. Xiao, R. Cheng, S. K. Kim, O. Tchernyshyov, O. A. Tretiakov, K.-W. Kim, and M. D. Stiles. This work was supported by the National Research Foundation of Korea (NRF) (2015M3D1A1070465).
\end{acknowledgments}


\begin{thebibliography}{99}
\bibitem{note} For a monolayer antiferromagnet, the initial N\'{e}el antiferromagnetic domain wall is broken when $J>4 \times 10^{11} \, \mathrm{A/m^2}$. The threshold current density to break the domain wall structure increases with the number of monolayers (not shown). The main conclusion of this work is not altered for multiple monolayers.
\end{thebibliography}
\end{document}